\documentstyle[12pt,epsf]{article}
\newcommand{\alt}{\mathrel{\raisebox{-.6ex}{$\stackrel{\textstyle<}{\sim}$}}}
\newcommand{\agt}{\mathrel{\raisebox{-.6ex}{$\stackrel{\textstyle>}{\sim}$}}}
\def\overlay#1#2{\ifmmode \setbox 0=\hbox {$#1$}\setbox 1=\hbox to\wd 0{\hss
$#2$\hss }\else \setbox 0=\hbox {#1}\setbox 1=\hbox to\wd 0{\hss #2\hss }\fi
#1\hskip -\wd 0\box 1}
\begin{document}

\font\fortssbx=cmssbx10 scaled \magstep2
\hbox to \hsize{
%\special{psfile=uwlogo.ps hscale=8000 vscale=8000 hoffset=-12 voffset=-2}
%\hskip.5in \raise.1in
\hbox{\fortssbx University of Wisconsin - Madison}
\hfill$\vcenter{\hbox{\bf MADPH-95-910}
                \hbox{\bf RAL-TR-95-069}
                \hbox{\bf IUHET-323}
                \hbox{November 1995}}$ }

\vspace{.75in}

\begin{center}
{\large\bf Renormalization group evolution of R-parity-violating Yukawa
couplings}\\[.6in]
V. Barger$^{(a)}$, M.S. Berger$^{(b)}$, R.J.N. Phillips$^{(c)}$, and T.
W\"ohrmann$^{(a)}$\\[.3in]
\it $^{(a)}$Physics Department, University of Wisconsin, Madison, WI 53706,
USA\\
$^{(b)}$Physics Department, Indiana University, Bloomington, IN 47405, USA\\
$^{(c)}$Rutherford Appleton Laboratory, Chilton, Didcot, Oxon OX11 0QX,
UK
\end{center}

\vspace{.75in}

\begin{abstract}\noindent
We study the evolution of $R$-parity-violating (RPV) couplings in the
minimum supersymmetric standard model, between the electroweak and
grand unification scales, assuming a family hierarchy for these
coupling strengths. Particular attention is given to solutions where
both the $R$-conserving and $R$-violating top quark Yukawa couplings
simultaneously approach infrared fixed points; these we analyse both
algebraically and with numerical solutions of the evolution equations
at one-loop level.   We identify constraints on these couplings at the
GUT scale,  arising from lower limits on the top quark mass. We show
that fixed points offer a new source of bounds on RPV couplings at the
electroweak scale.  We derive evolution equations for the CKM matrix,
and show that RPV couplings affect the scaling of the unitarity triangle.
The fixed-point behaviour is compatible with all present experimental
constraints.  However, fixed-point values of RPV top-quark couplings
would require the corresponding sleptons or squarks to have mass $\agt m_t$
to suppress strong new top decays to sparticles.

\end{abstract}

\thispagestyle{empty}

\newpage

\section{Introduction}

Supersymmetry is a very attractive extension of the Standard Model (SM), with
low-energy implications that are being actively pursued, both theoretically and
experimentally~\cite{susy1,susy2}. In the minimal supersymmetric extension of
the standard model (MSSM), with minimum new particle content, a discrete
symmetry ($R$-parity) is assumed to forbid rapid proton decay. In terms of
baryon number $B$, lepton number $L$ and spin $S$, the $R$-parity of a particle
is $R\equiv (-1)^{3B+L+2S}$, with value $R=+1$ for particles and $R=-1$ for
sparticles. An important consequence of $R$-conservation is that the lightest
sparticle is stable and is thus a candidate for cold dark matter. However,
since $R$-conservation is not theoretically motivated by any known principle,
the possibility of $R$-nonconservation deserves equally serious consideration.
In addition to the Yukawa superpotential in the MSSM
\begin{equation}\label{supmssm}
{\cal W} = ({\bf U })_{ab} H_2 Q_L^a \bar U_R^b + ({\bf D })_{ab} H_1 Q_L^a
\bar D_R^b + ({\bf E })_{ab} H_1 L_L^a \bar E_R^b,
\end{equation}
there are two classes of $R$-violating couplings in the MSSM
superpotential, allowed by supersymmetry and renormalizability~\cite{rpv}. The
superpotential terms for the first class violate lepton number $L$,
\begin{equation}
{\cal W} =
\lambda_{abc} L_L^a L_L^b \bar E_R^c + \lambda'_{abc} L_L^a Q_L^b
\bar D_R^c \,,
\end{equation}
while those of the second class violate baryon number $B$,
\begin{equation}
{\cal W} =
\lambda''_{abc} \bar D_R^a \bar D_R^b \bar U_R^c \,.
\end{equation}
Here $L, Q, \bar E, \bar D, \bar U$ stand for the doublet lepton, doublet
quark, singlet antilepton, singlet $d$-type antiquark, singlet $u$-type
antiquark superfields, respectively, and $a,b,c$ are generation indices.
The $ ({\bf U })_{ab}$, $ ({\bf D })_{ab}$ and  $ ({\bf E })_{ab}$ in
Eq.~(\ref{supmssm}) are the Yukawa coupling matrices. In our notation, the
superfields above are the weak interaction eigenstates, which might be expected
as the natural choice at the grand unified scale, rather than the
mass eigenstates. The Yukawa couplings
$\lambda_{abc}$ and $\lambda''_{abc}$ are antisymmetric in their first
two indices because of superfield antisymmetry.
These superpotential terms lead to the interaction lagrangians
\begin{eqnarray}
{\cal L} & = & \lambda_{abc} \{ \tilde \nu_{aL} \bar e_{cR} e_{bL} +
\tilde e_{bL} \bar e_{cR} \nu_{aL} + (\tilde e_{cR})^{\ast }
(\bar \nu_{aL})^c e_{bL} - ( a \leftrightarrow  b ) \} + h.c.
\end{eqnarray}
for the $\lambda$-terms, whereas the $\lambda'$-terms yield
\begin{eqnarray}
{\cal L} &=& \lambda'_{abc} \{\tilde \nu_{aL} \bar d_{cR} d_{bL} +
\tilde d_{bL} \bar d_{cR} \nu_{aL} + (\tilde d_{cR})^{\ast }
(\bar \nu_{aL})^c d_{bL} \nonumber\\
&&\qquad - \tilde e_{aL} \bar d_{cR} u_{bL} - \tilde u_{bL}
\bar d_{cR} e_{aL} - (\tilde d_{cR})^{\ast }
(\bar e_{aL})^c u_{bL} \} + h.c. \;,
\end{eqnarray}
with corresponding terms for each of these generations. In the case of a
$B$-violating superpotential the lagrangian reads
\begin{eqnarray}
{\cal L} & = &\lambda''_{abc} \{ u_c^c d_a^c \tilde d_{b}^{\ast } +
u_c^c \tilde d_{a}^{\ast } d_b^c + \tilde u_{c}^{\ast } d_a^c d_b^c \} +h.c.
\end{eqnarray}
To escape the proton-lifetime constraints, it is
sufficient that only one of these classes be absent or very highly suppressed.
Phenomenological studies of the consequences of $R$-parity violation (RPV) have
placed constraints on the various couplings $\lambda_{abc}, \lambda'_{abc},
\lambda''_{abc}$~\cite{bgh,const,tata,agashe,roy}, but considerable latitude
remains for RPV.

Studies of the renormalization group evolution equations (RGE), relating
couplings at the electroweak scale to their values at the grand unification
(GUT) scale, have led to new insights and constraints on the observable
low-energy parameters in the $R$-conserving scenario. It therefore seems
worthwhile to see what can be learned from similar studies of RPV scenarios. An
initial study of this type addressed the evolution of $\lambda''_{133}$ and
$\lambda''_{233}$ couplings~\cite{roy}. This was subsequently extended to
all the baryon-violating couplings $\lambda''_{ijk}$~\cite{goitysher}.
In the  present work we undertake a
somewhat more general study of the RGE for RPV interactions,
paying particular attention to solutions for which both the $R$-conserving
and $R$-violating top quark Yukawa couplings simultaneously approach
infrared fixed points.  Such fixed-point behaviour requires a
coupling $\lambda,\;\lambda'$, or $\lambda''$ to be of order unity at the
electroweak scale.  After our study was completed, a related work
on RGE for RPV couplings appeared~\cite{Herbi}, which however has a
different focus and is largely complementary to the present paper.

In the context of grand unified theories one is led to consider the
possible unification of RPV parameters.
If for example the RPV interactions arose from an SU(5)-invariant term,
then in fact the $L$-violating RPV couplings would be related to the
$B$-violating ones~\cite{sv} at the GUT scale.
We could then no longer set one or the other arbitrarily to
zero and the proton lifetime (which places very strong constraints on
products of $L$-violating and $B$-violating RPV couplings, typically
requiring products $\lambda' \lambda'' $ to be  smaller than
$5\times 10^{-17}$~\cite{sv}) would strongly constrain all types of
RPV couplings.   It can be argued that some products of $B$-violating
and $L$-violating couplings, containing several high-generation
indices, would not contribute directly to proton decay~\cite{carlson};
however,  proton decay would still be induced at the one-loop
level by flavor mixing~\cite{sv}, so in fact all RPV couplings
would have to be very small.
In such scenarios the fixed-point solutions for RPV couplings
would be excluded; our present studies therefore implicitly assume
that this kind of RPV unification does not occur.
Furthermore, since RPV unification is analogous to the popular
hypothesis of $\lambda _b=\lambda _\tau $ Yukawa unification, it
would appear somewhat inconsistent (though not completely unthinkable)
to assume one without the other.  Accordingly, in our present work, we
do not try to impose the additional constraint of
$\lambda _b=\lambda _\tau $ unification.

\section{Renormalization group equations and fixed points}

For any trilinear term in the superpotential $d_{abc} \Phi^a \Phi^b \Phi^c$
involving superfields $\Phi^a, \Phi^b, \Phi^c$, the evolution of the couplings
$d_{abc}$ with the scale $\mu$ is given by the RGE
\begin{equation}\label{dabcequ}
\mu {\partial\over\partial\mu} d_{abc} = \gamma_a^e d_{ebc} + \gamma_b^e
d_{aec} + \gamma_c^e d_{abe} \,,
\end{equation}
where the $\gamma_a^e$ are elements of the anomalous dimension matrix.
Table~I gives the anomalous dimensions for the superfields. The first column of
the table gives the results for the MSSM in matrix form; here {\bf U,\ D} and
{\bf E} are the matrices of Yukawa couplings to the up-quarks, down-quarks and
charged leptons, respectively, and a unit matrix is understood in front of the
terms involving SU(3), SU(2) and U(1) gauge couplings $g_3,\ g_2$ and $g_1$
and the terms with traces.  The second
column of Table~\ref{violating} gives the additions to the anomalous dimension
matrix due to $L$-violating terms $\lambda_{abc}$ and $\lambda'_{abc}$, while
the third column gives the corresponding additions due to $B$-violating
$\lambda''_{abc}$ terms. In our notation, an RPV-coupling with upper indices
is the complex conjugate of the same coupling with lower indices, e.g.
$\lambda ^{abc}=\lambda _{abc}^*$.

\begin{table}[t]
\caption{\label{violating}
$16\pi^2\gamma_{\phi_j}^{\phi_i}$ in the MSSM plus additional terms for lepton
or baryon number violating couplings, where $i$ and $j$ are flavor indices.}
\[\begin{array}{cccc}
\hline\hline
\phi_{i,j}& \rm MSSM& \mbox{Lepton \# Violation}& \mbox{Baryon \# Violation}\\
\hline
L_{i,j}& {\bf E E^\dagger}-{3\over2}g_2^2-{3\over10}g_1^2 &
\lambda_{iab}\lambda^{jab}+3\lambda'_{iab}\lambda'^{jab}& \mbox{---}\\
E_{i,j} & 2{\bf E^\dagger E}-{6\over5}g_1^2 & \lambda^{abi}\lambda_{abj}&
\mbox{---}\\
D_{i,j} & 2{\bf D^\dagger D}-{8\over3}g_3^2-{2\over15}g_1^2&
2\lambda'^{abi}\lambda'_{abj} & 2\lambda''^{iab}\lambda''_{jab}\\
U_{i,j} & 2{\bf U^\dagger U}-{8\over3}g_3^2-{8\over15}g_1^2& \mbox{---}&
\lambda''^{abi}\lambda''_{abj}\\
Q_{i,j} & {\bf UU^\dagger  + DD^\dagger }
-{8\over3}g_3^2-{3\over2}g_2^2-{1\over30}g_1^2 & \lambda'_{aib}\lambda'^{ajb}&
\mbox{---}\\
H_1 & {\rm Tr}({\bf EE^\dagger })+3{\rm Tr}({\bf DD^\dagger })
-{3\over2}g_2^2-{3\over10}g_1^2& \mbox{---} & \mbox{---}\\
H_2 & 3{\rm Tr}({\bf UU^\dagger })-{3\over2}g_2^2-{3\over10}g_1^2 & \mbox{---}&
\mbox{---}\\
\hline\hline
\end{array}\]
\end{table}

The evolution equations for the $R$-conserving Yukawa matrices ${\bf U},
{\bf D} , {\bf E }$ of Eq. (\ref{supmssm}) are obtained from
Eq.~(\ref{dabcequ}) with the index $c$ belonging to a Higgs field.
The general forms of the RGE are
\begin{eqnarray}
\mu {\partial\over\partial\mu} ({\bf U })_{ab} &=& ({\bf U })_{ib}
 \gamma_{Q_a}^{Q_i} + ({\bf U })_{ai}\gamma^{\bar U_i}_{\bar U_b} +
({\bf U })_{ab}\gamma_{H_2}^{H_2}
 \,, \label{rgegen1} \\
\mu {\partial\over\partial\mu} ({\bf D })_{ab} &=& ({\bf D })_{ib}
 \gamma_{Q_a}^{Q_i} + ({\bf D })_{ai}\gamma^{\bar D_i}_{\bar D_b} +
({\bf D })_{ab}\gamma_{H_1}^{H_1}
\,,\\
\mu {\partial\over\partial\mu} ({\bf E })_{ab} &=& ({\bf E })_{ib}
 \gamma_{L_a}^{L_i} + ({\bf E })_{ai}\gamma^{\bar E_i}_{\bar E_b} +
({\bf E })_{ab}\gamma_{H_1}^{H_1} \, . \label{rgegen3}
\end{eqnarray}
When we solve Eqs.~(\ref{rgegen1})--(\ref{rgegen3}) for the general
$R$-parity violating case,
we get additional contributions from Hermitian matrices
involving the RPV couplings that are analogous to combinations like
${\bf D^\dagger D}$ for the usual Yukawa matrices. For example the matrix
equation for the Yukawa matrices ${\bf U}$ and ${\bf D}$ become
\begin{eqnarray}
{{d{\bf U}}\over {dt}}&=&{1\over {16\pi ^2 }}
\Bigg[\Big [-{16\over3}\alpha_3-3\alpha_2-{13\over15}\alpha_1 \nonumber \\
&&{}+3{\bf UU^{\dagger }}+{\bf DD^{\dagger }}
+{\bf Tr}[3{\bf UU^{\dagger }}]+{\bf M}^{'(Q)}\Big ]
{\bf U}+{\bf UM}^{''(U)}\Bigg]\;,
\label{dUdt} \\
{{d{\bf D}}\over {dt}}&=&{1\over {16\pi ^2 }}
\Bigg[\Big [-{16\over3}\alpha_3-3\alpha_2-{7\over15}\alpha_1 \nonumber \\
&&{}+3{\bf DD^{\dagger }}+{\bf UU^{\dagger }}
+{\bf Tr}[3{\bf DD^{\dagger }}+{\bf EE^{\dagger }}]+{\bf M}^{'(Q)}\Big ]
{\bf D}+2{\bf DM}^{''(D)}+2{\bf DM}^{'(D)}\Bigg]\;,\qquad
\label{dDdt}
\end{eqnarray}
where ${\bf M}^{'(Q)}_{ij}\equiv \lambda'_{aib}\lambda'^{ajb}$,
${\bf M}^{'(D)}_{ij}\equiv \lambda'^{abi}\lambda'_{abj}$,
${\bf M}^{''(U)}_{ij}\equiv \lambda''^{abi}\lambda''_{abj}$ and
${\bf M}^{''(D)}_{ij}\equiv \lambda''^{iab}\lambda''_{jab}$
are the combinations of RPV couplings appearing in Table I.
The variable is
\begin{equation}
t=\ln (\mu/M_G)
\end{equation}
where $\mu$ is the running
mass scale and $M_G$ is the GUT unification mass.

The gauge couplings are not affected by the
presence of $R$-violating couplings at the one-loop level.

The third generation Yukawa couplings are dominant, so if we retain
in the anomalous dimensions only the (3,3) elements $\lambda_t, \lambda_b,
\lambda_\tau$ in $\bf U, D, E$, setting all other elements to zero,
Eqs. (\ref{rgegen1})--(\ref{rgegen3}) read
\begin{eqnarray}
\mu {\partial\over\partial\mu} \lambda_t &=& \lambda_t
\left[ \gamma_{Q_3}^{Q_3} + \gamma^{\bar U_3}_{\bar U_3} + \gamma_{H_2}^{H_2}
\right] \,,\\
\mu{\partial\over\partial\mu}\lambda_b &=& \lambda_b
\left[\gamma_{Q_3}^{Q_3}+\gamma_{\bar D_3}^{\bar D_3}+\gamma_{H_1}^{H_1}\right]
\,,\\
\mu{\partial\over\partial\mu} \lambda_\tau &=& \lambda_\tau
\left[\gamma_{L_3}^{L_3}+\gamma_{\bar E_3}^{\bar
E_3}+\gamma_{H_1}^{H_1}\right]\,,
\end{eqnarray}

Since there are 36 independent RPV couplings $\lambda_{abc},\lambda'_{abc}$ in
the $L$-violating sector (9~independent couplings $\lambda''_{abc}$ in the
$B$-violating sector) to be added to the three dominant $R$-conserving Higgs
couplings $\lambda_t,\lambda_b,\lambda_\tau$, we would have to consider 39 (12)
coupled non-linear evolution equations, in general. Some further radical
simplifications in the RPV sector
are clearly needed to make the system of equations tractable.

It is plausible that there may exist a generational hierarchy among the RPV
couplings, analogous to that of the conventional Higgs couplings; indeed, the
RPV couplings to higher generations evolve more strongly due to larger Higgs
couplings in their RGE, and hence have the potential to take larger values than
RPV couplings to lower generations. Thus we consider retaining only the
couplings $\lambda_{233}$ and $\lambda'_{333}$, or $\lambda''_{233}$,
neglecting all others. This restriction is also motivated by the fact
that the experimental upper limits are stronger for
the couplings with lower indices.

To simplify the form of the RGE, we adopt the following notation:
\[ Y_i = {1\over4\pi}\lambda_i^2\ (i=t,b,\tau),\quad
Y''={1\over4\pi}\lambda''^2_{233},\quad
Y' = {1\over4\pi}\lambda'^2_{333},\quad
Y = {1\over4\pi}\lambda^2_{233}.\]
The one-loop RGE then take the following forms,
where $\alpha_i = {1\over4\pi}g_i^2$,
\begin{eqnarray}
{d\alpha_i\over dt} &=& {1\over {2\pi}}
b_i\alpha_i^2 \;, \qquad b_i=\left \{ 33/5,1,-3\right \} \\
{dY_t\over dt} &=& {1\over {2\pi}}
Y_t \left(6Y_t+Y_b+Y'+2Y''
\textstyle-{16\over3}\alpha_3-3\alpha_2-{13\over15}\alpha_1\right)\\
{dY_b\over dt} &=& {1\over {2\pi}}
Y_b \left(Y_t+6Y_b+Y_\tau+3Y'+2Y''
\textstyle-{16\over3}\alpha_3-3\alpha_2-{7\over15}\alpha_1\right)\\
{dY_\tau\over dt} &=& {1\over {2\pi}}
Y_\tau \left(3Y_b+4Y_\tau+3Y+3Y'
\textstyle-3\alpha_2-{9\over5}\alpha_1\right)\\
{dY\over dt} &=& {1\over {2\pi}}
Y\left(3Y_\tau+3Y+3Y'
\textstyle-3\alpha_2-{9\over5}\alpha_1\right)\\
{dY'\over dt} &=& {1\over {2\pi}}
Y'\left(Y_t+3Y_b+Y_\tau+Y+6Y'
\textstyle-{16\over3}\alpha_3-3\alpha_2-{7\over15}\alpha_1\right)\\
{dY''\over dt} &=& {1\over {2\pi}}
Y''\left(2Y_t+2Y_b+6Y''
\textstyle-8\alpha_3-{4\over5}\alpha_1\right)
\end{eqnarray}
Here it is understood that one takes {\em either} $Y=Y'=0$ {\em or} $Y''=0$.

An extremely interesting possibility in the RGE is that $Y_t$ is large at the
GUT scale and consequently is driven toward a fixed point at the electroweak
scale~\cite{ross,bbo}. In particular, in the MSSM $\lambda_t\to 1.1$ as $\mu\to
m_t$; since $\lambda_t(m_t) = \sqrt2m_t(m_t)/(v\sin\beta)$, this leads to the
relation, for low $\tan\beta$~\cite{bbo}
\begin{equation}
m_t({\rm pole}) = (200\rm\ GeV)\sin\beta \,,
\end{equation}
where $\tan\beta=v_2/v_1$ is the ratio of the Higgs vevs and $m_t$(pole) is the
mass at the $t$-propagator pole. It is interesting to examine the impact
of RPV couplings on this fixed-point result~\cite{roy}.

\subsection{ $\lambda_t $ fixed point in the MSSM }

We first review the $\lambda_t$ fixed-point behavior in the MSSM limit, where
RPV couplings are neglected. Setting $dY_t/dt\simeq0$ at $\mu\simeq m_t$ gives
the fixed-point condition
\begin{equation}
6Y_t+Y_b = \textstyle{16\over3}\alpha_3+3\alpha_2+{13\over15}\alpha_1 \,.
\end{equation}
The $\lambda_t$ and $\lambda_b$ couplings at $\mu=m_t$ are related to the
running masses
\begin{equation}
\lambda_t(m_t) = {\sqrt2 m_t(m_t)\over v\sin\beta}\,, \qquad
\lambda_b(m_t) = {\sqrt2 m_b(m_b)\over \eta_b v\cos\beta} \,,
\end{equation}
with $v=\left(\sqrt2 \, G_F\right)^{-1/2} = 246$~GeV.
Here $\eta_b$ gives the QCD/QED running of $m_b(\mu)$ between $\mu=m_b$ and
$\mu=m_t$; $\eta_b\simeq1.5$ for $\alpha_s(m_t)\simeq0.10$~\cite{bbo}. Thus we
can express $\lambda_b(m_t)$ in terms of $\lambda_t(m_t)$, $\tan\beta$ and the
known running masses:
\begin{equation}
\lambda_b(m_t) = {m_b(m_b)\over m_t(m_t)} {\tan\beta\over \eta_b}
\lambda_t(m_t) \simeq 0.017 \tan\beta \; \lambda_t(m_t) \,,
\end{equation}
taking $m_b(m_b)=4.25$~GeV, $m_t(m_t)=167$~GeV, and hence
\begin{equation}
Y_b(m_t) \simeq 3\times10^{-4}\tan^2\!\beta \; Y_t(m_t) \,.
\end{equation}
For small or moderate values $\tan\beta\alt20$, we obtain
$Y_b/(6Y_t)<0.02$ so we can safely neglect the $Y_b$ contribution.
In this case, taking the approximate values
\begin{equation}
\alpha_3=1/10 ,\quad \alpha_2=1/30,\quad \alpha_1=1/58 \quad{\rm at}\ \mu=m_t,
\end{equation}
we find the numerical value
\begin{equation}
Y_t(m_t) =  0.108,  \qquad  \lambda_t(m_t)=1.16\, , \\
\end{equation}
For large $\tan\beta\sim m_t/m_b$, we can express the
$\lambda_t$ fixed-point relation as
\begin{equation}
Y_t(m_t) = {\lambda_t^2(m_t)\over 4\pi} = \left(\textstyle{8\over9}\alpha_3
+{1\over2}\alpha_2 + {13\over90}\alpha_1\right) \Big/
\left(1+5\times10^{-5}\tan^2\beta\right)\,.
\end{equation}

\subsection{ $\lambda'' $, $\lambda_t $ simultaneous fixed points }

Next we consider the $B$-violating scenario with $Y=Y'=0$ and $Y''$ non-zero,
investigating the possibility that fixed-point limits are approached for both
$Y_t$ and $Y''$ couplings, as found numerically in Ref.~\cite{roy} (note that
these authors use a different definition of $\lambda''_{abc}$). This
requires $dY_t/dt\simeq0$ and $dY''/dt\simeq0$ at $\mu\simeq m_t$, giving the
conditions
\begin{eqnarray}
6Y_t+Y_b+2Y'' \textstyle -{16\over3}\alpha_3 - 3\alpha_2 - {13\over15}\alpha_1
&\simeq& 0\,,\\
2Y_t+2Y_b+6Y''\textstyle -8\alpha_3 \phantom{{}-3\alpha_2} - {4\over5}\alpha_1
&\simeq& 0\,. \label{fixed}
\end{eqnarray}
Taking linear combinations to solve for $Y_t$ and $Y''$ we obtain (with $Y_b\ll
Y_t$)
\begin{eqnarray}&&
Y_t\ \simeq\textstyle{1\over16} \left(8\alpha_3+9\alpha_2+
{9\over5}\alpha_1\right)\ \ \, \simeq 0.071 \,,\qquad\;\ \lambda_t \simeq
0.94\,, \label{fixed1}\\
&&Y'' \simeq \textstyle {1\over16}\left( {56\over3}\alpha_3 - 3\alpha_2 +
{23\over15}\alpha_1\right) \simeq 0.112 \,,\qquad \lambda''_{233}
\simeq 1.18\,,
\label{fixed2}
\end{eqnarray}
showing a considerable downward displacement in $\lambda_t$ due to
$\lambda''_{233}$. Such a large value of $\lambda''_{233}$ would imply
substantial $t\to b\tilde s, s\tilde b$ decay, if kinematically allowed.

If both $\lambda_t$ and $\lambda''_{233}$ fixed points are realized as above,
then the predicted physical top quark mass is
\begin{equation}
m_t(\rm pole) \simeq (150~GeV)\ \sin\beta \,.
\end{equation}
Even for moderate values of $\tan \beta \, ( \tan \beta > 5 )$ one has
$\sin\beta \simeq 1 \, ( \sin \beta > 0.98 )$.
This prediction is at the lower end of the present data~\cite{cdf,dzero}:
\begin{equation}
m_t = 176\pm 8\pm 10~{\rm GeV\quad (CDF)} \,, \qquad m_t = 199^{+10}_{-21} \pm
22~\rm GeV\quad(D0) \,.
\end{equation}
When the data become more precise, the fixed-point possibility for
$\lambda''_{233}$ could be excluded, if the measured central value of $m_t$ is
unchanged.

One can also consider the case of large $\tan \beta$ where the coupling $Y_b$
is non-negligible, and in fact may be near its own fixed point. In that case
we add another equation, $dY_b/dt\simeq0$, to those above. This gives
\begin{eqnarray}
Y_t+6Y_b+Y_\tau+2Y''
\textstyle-{16\over3}\alpha_3-3\alpha_2-{7\over15}\alpha_1 \simeq 0\;.
\end{eqnarray}
A new coupling $Y_\tau$ enters here, but it can be related to $Y _b$
since
\begin{equation}
\lambda_\tau(m_t) = {\sqrt2 m_\tau(m_t)\over \eta_\tau v\cos\beta} \,,
\end{equation}
and hence
\begin{eqnarray}
\lambda_\tau(m_t) = {m_\tau(m_\tau)\over m_b(m_b)} {\eta_b\over\eta_\tau}
\lambda_b(m_t) = 0.6\lambda_b(m_t)\;,\quad
Y_\tau(m_t) = 0.4 Y_b(m_t)\;,
\end{eqnarray}
by arguments similar to those above relating $\lambda_b(m_t)$ to
$\lambda_t(m_t)$.  Then we have three simultaneous equations
in three unknowns, that give the solutions
\begin{eqnarray}&&
Y_t\ \simeq 0.067 \,,\qquad \lambda_t \simeq 0.92 \,,
\\
&&Y_b\ \simeq 0.061 \,,\qquad \lambda_b \simeq 0.88 \,,
\\
&&Y'' \simeq 0.092 \,,\qquad \lambda''_{233} \simeq 1.08\,.
\end{eqnarray}

\subsection{ $\lambda $, $\lambda'$, $\lambda_t $ simultaneous fixed points }
\label{subsllp}

If instead fixed points should occur simultaneously for $Y_t$ and $Y'$, the
conditions at $\mu\simeq m_t$, found from $dY_t/dt\simeq0$ and $dY'/dt\simeq
0$, are
\begin{eqnarray}
Y_t &=& \textstyle {1\over35} \left[ {80\over3}\alpha_3 + 15\alpha_2 +
{71\over15}\alpha_1 - 3Y_b + Y_\tau + Y \right] \,,\\
Y' &=& \textstyle {1\over35} \left[ {80\over3}\alpha_3 + 15\alpha_2
+{29\over15}\alpha_1 - 17Y_b -6Y_\tau -6Y \right] \,.
\end{eqnarray}
If $Y$ is small and we also neglect $Y_b$ and $Y_\tau$ (e.g. assuming
small $\tan\beta$), then $Y_t$ and $Y'$ approach almost the same fixed-point
value
\begin{eqnarray}
\lambda_t(m_t) \simeq \lambda'_{333} \simeq 1.07 \,.\label{fixed3}
\end{eqnarray}
Alternatively, if $Y_b$ is large, all three couplings $Y_t$, $Y_b$ and
$Y'$ can approach fixed points; the solution of the corresponding
three equations gives
\begin{eqnarray}
\lambda_t(m_t) \simeq 1.05\,,\quad
\lambda_b \simeq\lambda'_{333} \simeq 0.86 \,.\label{fixed4}
\end{eqnarray}
In both the above cases $\lambda_t(m_t)$ is only slightly displaced
below the MSSM value, while $\lambda'_{333}$ has quite a large value.
The latter would imply substantial $t\to b\tilde{\bar\tau},
\bar\tau\tilde b$ decays, if kinematically allowed;
the $t\to b\tilde{\bar\tau}$ mode is more likely,
since $\tilde{\bar\tau}$ is usually expected to be lighter than
$\tilde b$, and we discuss its implications later.

%It is not possible for $Y_t$, $Y$ and $Y'$ all to approach fixed
%points simultaneously, because the corresponding conditions would
%require negative $Y$.  However,
If $Y'$ is negligible, $Y_t$ and $Y$
can approach fixed points simultaneously; in this case the two
conditions essentially decouple, giving the MSSM result for $Y_t$.
If $Y_b$ and $Y_{\tau}$ are negligible, the solution is
\begin{eqnarray}
\lambda_t(m_t) \simeq 1.16,\quad
\lambda_{233}\simeq 0.74 \,,\label{fixed5}
\end{eqnarray}
but if $Y_b$ too is large and approaches its fixed point,
the three corresponding conditions give
\begin{eqnarray}
\lambda_t(m_t) \simeq 1.09\,,\quad
\lambda_b \simeq 1.04\,,\quad\lambda_{233} \simeq 0.40 \,.\label{fixed6}
\end{eqnarray}

\subsection{ CKM evolution }

The presence of non-zero RPV couplings can also change the
evolution of CKM mixing angles. This has interesting implications for the
prediction of fermion mixings at the electroweak scale
from an ansatz for Yukawa matrices at the GUT scale.
In a model such as the MSSM (or the SM) with no RPV terms, the
evolution of the CKM angles at the one-loop level comes entirely from the
Yukawa matrix terms in the anomalous dimension $\gamma ^{Q_i}_{Q_j}$.
The Yukawa matrices ${\bf U}$ and ${\bf D}$ can be diagonalized by
bi-unitary transformations
\begin{eqnarray}
{\bf U^{diag}}&=&V_U^L{\bf U}V_U^{R\dagger} \;, \\
{\bf D^{diag}}&=&V_D^L{\bf D}V_D^{R\dagger} \;.
\end{eqnarray}
The CKM matrix is then given by
\begin{equation}
V\equiv V_U^LV_D^{L\dagger } \;.
\end{equation}
In the presence of RPV there are additional contributions
to the anomalous dimensions and hence to the
CKM RGE's. Consider for example the case in which only the $\lambda^{''}$
couplings are nonzero, for which there are new contributions
${\bf M}^{''(U)}_{ij}$ and
${\bf M}^{''(D)}_{ij}$ to the RGE's
as defined following Eq.~(\ref{dDdt}). The RPV contributions to the RGE's
can be diagonalized by
\begin{eqnarray}
{\bf M}^{''(U),{\bf diag}}&=&V_{(U)}^R
{\bf M}^{''(U)}V_{(U)}^{R\dagger}\equiv
\left \{\lambda _u^{''2},\lambda _c^{''2},\lambda _t^{''2}\right \} \;, \\
{\bf M}^{''(D),{\bf diag}}&=&V_{(D)}^R
{\bf M}^{''(D)}V_{(D)}^{R\dagger}\equiv
\left \{\lambda _d^{''2},\lambda _s^{''2},\lambda _b^{''2}\right \} \;,
\end{eqnarray}
for which new matrices
\begin{eqnarray}
&&V^{(U)}\equiv V_U^RV_{(U)}^{R\dagger} \;, \label{VU} \\
&&V^{(D)}\equiv V_D^RV_{(D)}^{R\dagger} \;, \label{VD}
\end{eqnarray}
can be defined.
We find the RGE's take the form
\begin{eqnarray}
{{dV_{i\alpha }}\over {dt}}&=&{1 \over {16\pi ^2}}\left [
\sum _{\beta,j\ne i}
{{\lambda _i^2+\lambda _j^2}\over {\lambda _i^2-\lambda _j^2}}
\lambda _{\beta}^2
V_{i\beta }V_{j\beta }^*V_{j\alpha }+\sum _{j,\beta \ne \alpha}
{{\lambda _{\alpha }^2+\lambda _{\beta }^2}\over {\lambda _{\alpha }^2
-\lambda _{\beta }^2}}\lambda _j^2
V_{j\beta }^*V_{j\alpha }V_{i\beta }\right . \nonumber \\
&&\quad +\left .\sum _{k,j\ne i}
{{\lambda _i\lambda _j}\over {\lambda _i^2-\lambda _j^2}}
\lambda _{k}^{''2}
V^{(U)}_{ik }V^{(U)*}_{jk }V_{j\alpha }+\sum _{\gamma,\beta \ne \alpha}
{{2\lambda _{\alpha }\lambda _{\beta }}\over {\lambda _{\alpha }^2
-\lambda _{\beta }^2}}\lambda _\gamma^{''2}
V_{\gamma\beta }^{(D)*}V_{\gamma\alpha }^{(D)}V_{i\beta }\right ] .
\end{eqnarray}
where $i,j,k=u,c,t$ and $\alpha,\beta,\gamma=d,s,b$.
One observes that generally there is a contribution to the evolution of
the CKM matrix from the RPV sector.

Assuming, as we do, that only the RPV couplings $\lambda_{233}$,
$\lambda_{333}^{\prime }$ or $\lambda_{233}^{\prime\prime }$ are non-zero,
the off-diagonal elements of the matrices defined in Eqs.~(\ref{VU}) and
(\ref{VD}) vanish. Then the one-loop RGEs
for mixing angles and the $CP$-violation parameter
$J={\rm Im}(V_{ud}^{}V_{cs}^{}V_{us}^*V_{cd}^*)$
have the same forms as in the MSSM, namely~\cite{bbo2}
\begin{eqnarray}
{{dW}\over {dt}}&=&{{W}\over {16\pi^2}}\left (\lambda_t^2+\lambda_b^2
\right )\;, \label{ckm}
\end{eqnarray}
where $W=|V_{ub}|^2$, $|V_{cb}|^2$, $|V_{td}|^2$, $|V_{ts}|^2$ or $J$.
Nevertheless the evolution of CKM angles  differs from the
MSSM because the evolution of the Yukawa couplings on the right
hand side is altered by the RPV couplings.

\section{Numerical RGE Studies}

In the previous section, we identified the quasi-infrared fixed points
that can be determined through the algebraic solutions to the RGE equations.
The one-loop RGEs form  a set of coupled first-order differential
equations that must be solved numerically.

Figure 1 shows the fixed point behaviour of each of the three RPV couplings
considered in this paper
($ \lambda^{\prime \prime }_{233}, \lambda^{\prime }_{333}, \lambda_{233}$)
along with the corresponding fixed point behaviour for $\lambda_t $,
assuming that $\tan\beta$ is small and hence
$\lambda_b$ and $\lambda_\tau$ are negligible.
It can be seen that for all $\lambda\agt 1$ at the GUT scale, the
respective Yukawa coupling approaches its fixed point at the electroweak scale.
These infrared fixed points provide the theoretical upper limits for the
RPV-Yukawa couplings at the electroweak scale summarized in Table~II.
The numerical evolution of the fixed points approaches but does not
exactly reproduce the approximate analytical values
Eqs.~(\ref{fixed1}),~(\ref{fixed2}),~(\ref{fixed3}) and (\ref{fixed5}).

\begin{table}[t]
\caption{ Fixed points for the different Yukawa couplings $\lambda$ in
different models for i)~$\tan\beta\protect\alt30$ and ii)~$\tan\beta\sim
m_t/m_b$.
In the case of large $\tan \beta $, $\lambda_b $ also reaches
a fixed point.}
\centering
\tabcolsep=1em
\begin{tabular}{@{~}rclllll}
\hline \hline
& Model & $\lambda_t $ & $\lambda_b $ & $ \lambda_{233} $ & $
\lambda_{333}^{\prime }$ & $\lambda_{233}^{\prime \prime } $  \\ \hline
i)& MSSM & 1.06 & -- & -- & -- & -- \\
& Lepton \# Violation ($\lambda \gg \lambda' $)& 1.06 & -- & 1.04 & -- & -- \\
& Lepton \# Violation ($\lambda' \gg \lambda $)& 0.99 & -- & -- & 0.97 & -- \\
& Baryon \# Violation & 0.90 & -- & -- & -- & 1.02 \\
\hline
ii)& MSSM & 1.00 & 0.92 & -- & -- & -- \\
&Lepton \# Violation ($\lambda \gg \lambda' $)& 0.99 & 0.98 & 1.04 & -- & -- \\
&Lepton \# Violation ($\lambda' \gg \lambda $)& 0.96 & 0.81 & -- & 0.80 & -- \\
&Baryon \# Violation & 0.87 & 0.85 & -- & -- & 0.92 \\
\hline \hline
\end{tabular}
\end{table}

We obtain additional restrictions on the RPV couplings from the
experimental lower bound on $m_t$ (that we take to be $m_t > 150$~GeV
\cite {cdf,dzero}). These additional limits are shown in Fig.~2;
the dark shaded region is excluded in all types of
models only by assuming this lower bound on the top mass.

One might hope that RPV interactions could help to explain the
measured value of
$R_b=\Gamma (Z\to b\bar b)/\Gamma (Z \to {\rm hadrons})$, which
differs from the SM prediction by over three standard deviations.
However, while their contributions can have either sign, the RPV
couplings must be significantly above their fixed-point values
to explain the full discrepancy~\cite{const}. In the case of lepton RPV the
bounds on the leptonic partial widths are always strong enough to prevent
RPV couplings from taking such large values.

Next we address the question, whether RPV couplings will
significantly change the relation between electroweak scale
and GUT scale values of the off-diagonal terms of the CKM matrix.
When the masses and mixings of the CKM matrix satisfy a hierarchy,
these relations are given by
\[      W(\mu )= W(\mbox{GUT}) S(\mu ),   \]
where $W$ is a CKM matrix element connecting the
third generation to one of the lighter generations, and
$S$ is a scaling factor~\cite{bbo2}.
The other CKM elements do not change with scale to leading order
in the hierarchy. The scaling factor $S(\mu )$ is determined by integrating
Eq.~(\ref{ckm}) together with the other RGEs. In Fig.~3 we show the
dependence of the scaling factor $S$ on the GUT-scale RPV couplings
$\lambda_{233}, \lambda_{333}^{\prime }$ and $ \lambda_{233}^{\prime \prime }$
respectively.

\begin{center}
\epsfxsize=5.2in
\hspace*{0in}
\epsffile{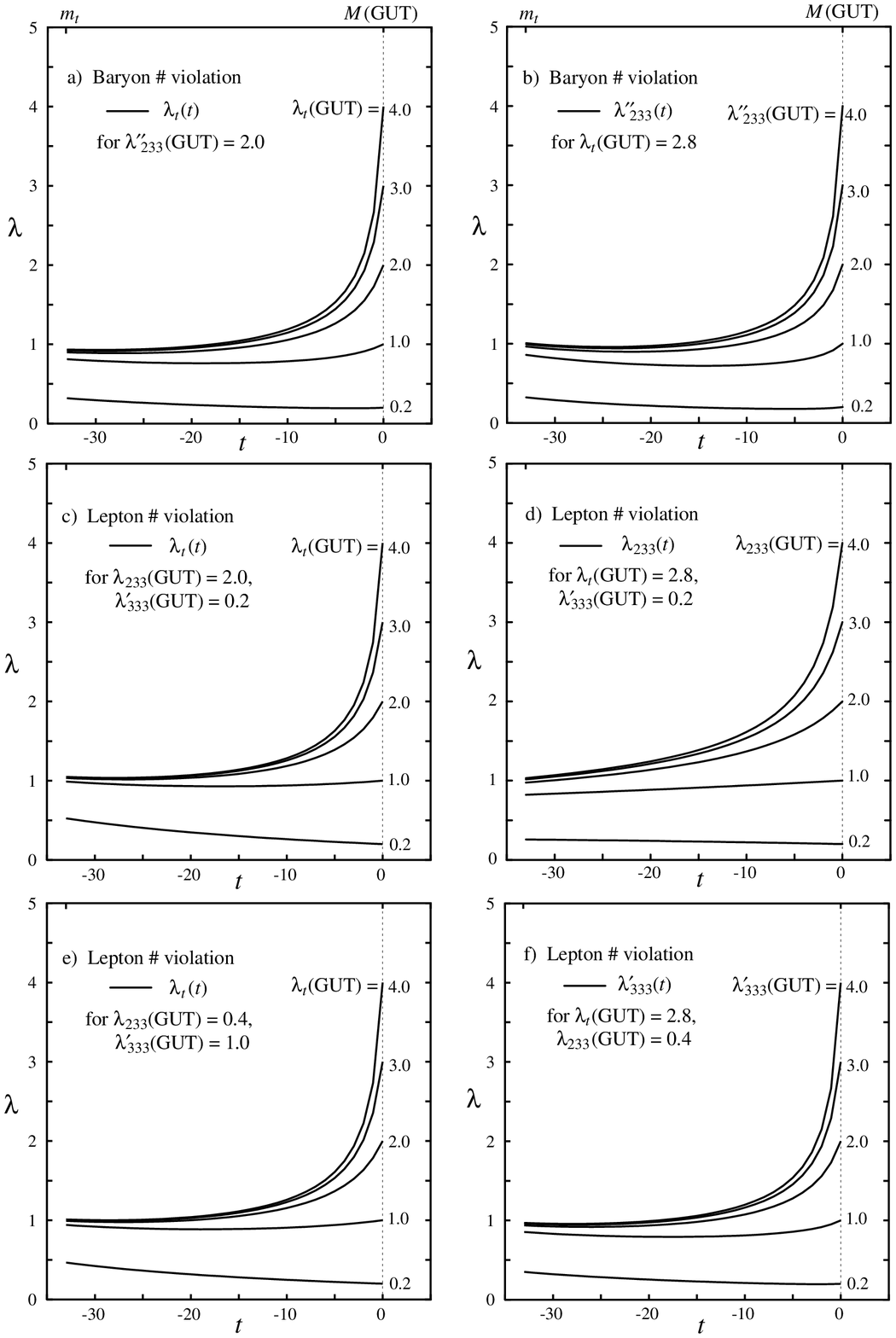}

\smallskip
\parbox{5.5in}{\small  Fig.~1. Couplings $\lambda$ as a function of the energy
scale $t$ for $\lambda_t $ in (a) baryon number RPV, (c)~lepton number RPV with
$\lambda_{233} \gg \lambda^{\prime}_{333}$ and (e)~lepton number RPV with
$\lambda^{\prime}_{333}\gg \lambda_{233} $ for
different starting points at the GUT scale ($t=0$). Panels~(b), (d) and (f)
show the same for $ \lambda^{\prime \prime}_{233}$, $\lambda_{233} \,
(\lambda_{233} \gg \lambda^{\prime}_{333})$ and $ \lambda^{\prime}_{333}
\, (\lambda^{\prime}_{333}\gg \lambda_{233})$ respectively.
Here $t\simeq -33$ represents the electroweak scale,
where these couplings reach their fixed points.}
\end{center}

\begin{center}
\epsfxsize=5.4in
\hspace*{0in}
\epsffile{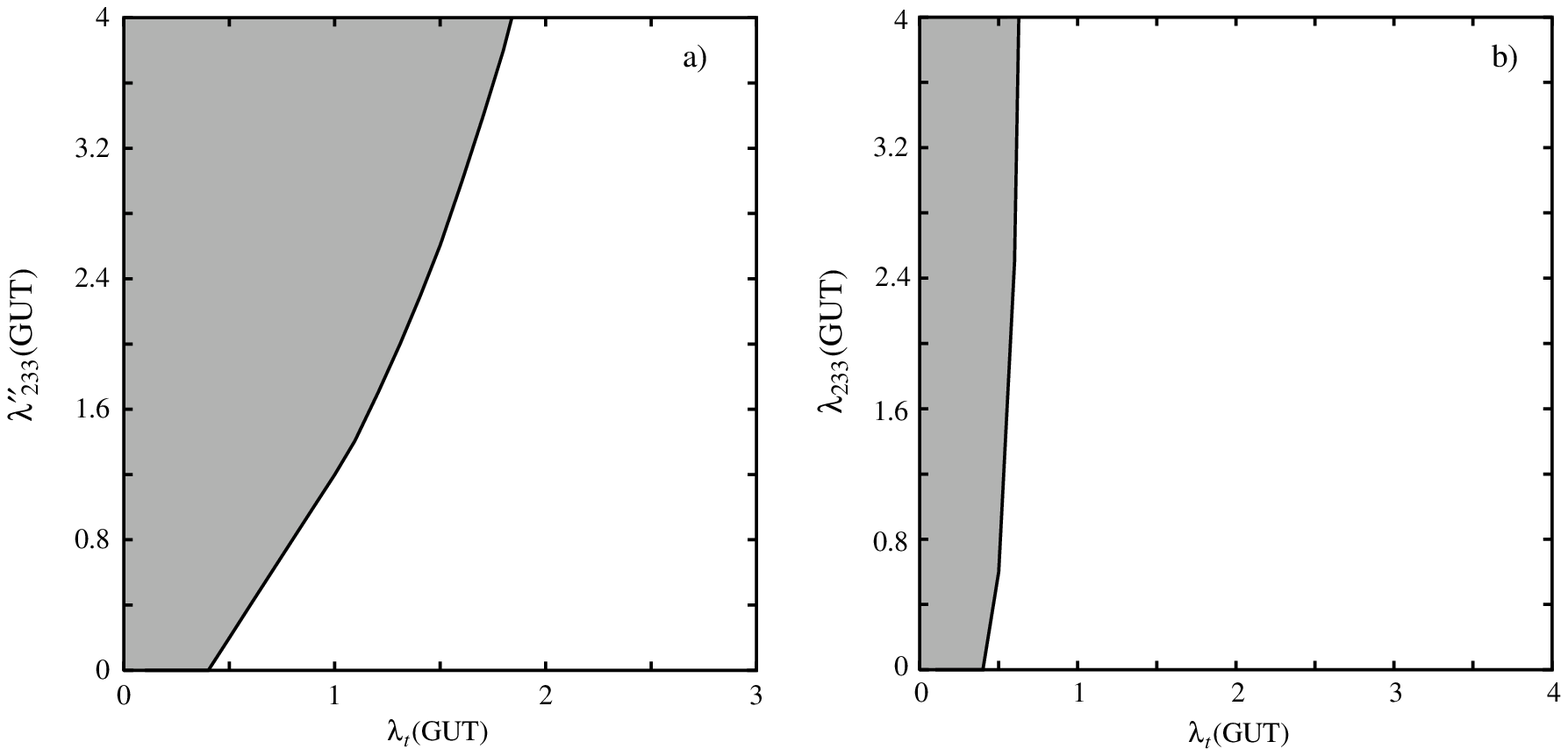}

\medskip
\parbox{5.5in}{\small Fig.~2. Excluded regions in the (a)
$\lambda_{t}\mbox{(GUT)}, \lambda_{233}^{\prime \prime} \mbox{(GUT)}$
plane and (b)~$\lambda_{t}\mbox{(GUT)}$, $\lambda_{233}\mbox{(GUT)}$
($\lambda_{233}\mbox{(GUT)}=2\lambda_{333}^\prime\mbox{(GUT)}$) plane
obtained from $ m_t > 150$~GeV. }
\end{center}

\begin{center}
\epsfxsize=5.2in
\hspace*{0in}
\epsffile{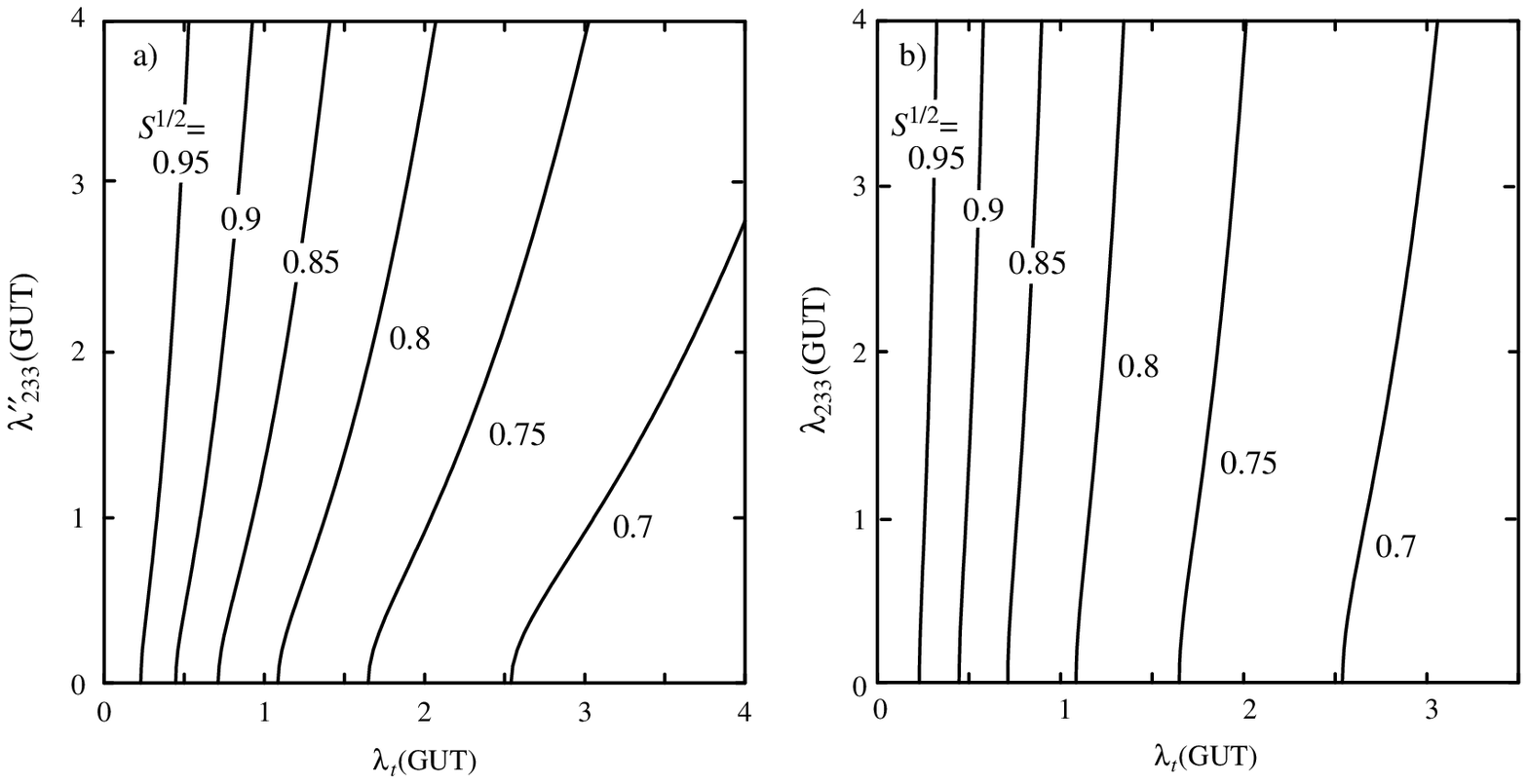}

\medskip
\parbox{5.5in}{\small \small Fig.~3.  Contours of constant $S^{1/2}$
for different values of (a) $\lambda_{233}^{\prime \prime} \mbox{(GUT)}$
and $\lambda_{t}\mbox{(GUT)}$ (baryon number violation) and (b)
$\lambda_{233}\mbox{(GUT)}=2\lambda_{333}^\prime\mbox{(GUT)}$ and
$\lambda_{t}\mbox{(GUT)}$ (lepton number violation).}
\end{center}

\section{RPV decays of the top quark}

The RPV couplings $\lambda^{\prime\prime}_{233}$ and $\lambda^{\prime}_{333}$
give rise to new decay modes of the top quark~\cite{dp}, if the
necessary
squark or slepton masses are small enough.

The $L$-violating coupling $\lambda^{\prime}_{333}$ leads to
$t_R\to {b_R}{\tilde{\bar {\tau}}}_R, \tilde{b}_R \bar{\tau}_R$ decays,
with partial widths~\cite{dp}
\begin{eqnarray}
\Gamma(t\to b\tilde{\bar\tau})&=&
{(\lambda^\prime_{333})^2\over 32\pi}\;
   m_t \; (1-m_{\tilde\tau}^2/m_t^2)^2 \; ,\\
\Gamma(t\to\tilde b\bar\tau)  &=&
{(\lambda^\prime_{333})^2\over 32\pi}\;
   m_t \; (1-m_{\tilde b}^2/m_t^2)^2 \; ,
\label{tbtau}
\end{eqnarray}
neglecting $m_b$ and $m_\tau$.  The former mode is more likely to be
accessible, since sleptons are expected to be lighter than squarks.
Since the SM top decay has partial width
\begin{equation}
\Gamma(t\to bW) = {G_Fm_t^3|V_{tb}|^2\over 8\pi\sqrt 2}\;
   (1-M_W^2/m_t^2)^2 \;(1+2M_W^2/m_t^2)\; ,
\label{tbw}
\end{equation}
the ratio of RPV to SM decays would be typically
\begin{equation}
\Gamma(t\to b\tilde {\tau}^+)/\Gamma(t\to bW^+)
\simeq 0.70\;(\lambda^\prime_{333})^2 \quad
 (for \; m_{\tilde\tau}\simeq M_W)\;.
\label{rpvsm1}
\end{equation}
It is natural to assume that $\tilde\tau$ would decay mostly to
$\tau$ plus the lightest neutralino $\chi^0_1$ (which is also
probably the lightest sparticle), followed by the RPV decay
$\chi^0_1\to b\bar b\nu_\tau(\bar\nu_\tau)$,
with a short lifetime~\cite{sd}
\begin{equation}
\tau (\chi^0_1\to b\bar b\nu_\tau, b\bar b\bar\nu_\tau)
\sim3\times 10^{-21}\;{\rm sec}\; (m_{\tilde b}/m_\chi)^4\;(100\;
{\rm  GeV}/m_\chi)/ (\lambda^\prime_{333})^2\; ,
\label{chilife}
\end{equation}
giving altogether
\begin{equation}
t\to b\tilde{\tau}^+\to b\tau\chi^0_1
                    \to bb\bar b\tau^+\nu_\tau(\bar\nu_\tau).
\label{tbbbtn}
\end{equation}
This mode could in principle be identified experimentally,
e.g.\ by exploiting the large number of potentially taggable $b$-jets
and the presence of a tau. However, it would not be
readily confused with the SM decay modes $t\to bW^+\to bq\bar q^\prime,
b\ell\nu$, ($\ell=e,\mu$), that form the basis of the presently
detected $p\bar p\to t\bar tX$ signals in the $(W\to\ell\nu)+4jet$
and dilepton channels (neglecting leptons from $\tau\to\ell\nu\nu$ that
suffer from a small branching fraction and a soft spectrum).
On the contrary, the RPV mode would deplete the SM signals by competition.
With $m_{\tilde\tau}\sim M_W$, fixed-point values
$\lambda^\prime_{333}\simeq 0.9$ (Fig.1) would suppress the SM signal
rate by a factor $(1+ 0.70(\lambda^\prime_{333})^2)^{-2}\simeq 0.4$, in
contradiction to experiment where $p\bar p\to t\bar tX\to b\bar bWWX$
signals tend if anything to exceed SM expectations~\cite{cdf,dzero}.
We conclude that either the fixed-point value is not approached
or the $\tilde\tau$ mass is higher and reduces the RPV effect
(e.g. $m_{\tilde\tau}=150$ GeV with $\lambda^\prime_{333}=0.9$
would suppress the SM signal rate by 0.88 instead).
Note that our discussion hinges on the fact that the RPV decays of
present interest would {\it not} contribute to SM top signals;
it is quite different from the approach of Ref.~\cite{agashe}, which
considers RPV couplings that would give hard electrons or muons
and contribute in conventional top searches.

Similarly, the $B$-violating coupling $\lambda^{\prime\prime}_{233}$
leads to
$t_R\to\bar{b}_R\tilde{\bar{s}}_R, \tilde{\bar{b}}_R\bar{s}_R$ decays,
with partial widths
\begin{equation}
\Gamma(t\to \bar b\tilde{\bar s})=\Gamma(t\to\tilde{\bar b}\bar s)\\
 = {(\lambda^{\prime\prime}_{233})^2\over 32\pi}\;
   m_t \; (1-m_{\tilde q}^2/m_t^2)^2 \; ,
\label{tbs}
\end{equation}
neglecting $m_b$ and $m_s$ and assuming a common squark mass
$m_{\tilde b}=m_{\tilde s}=m_{\tilde q}$.  If the squarks were
no heavier than 150 GeV, say, the ratio of RPV to SM decays would be
\begin{equation}
\Gamma(t\to \bar b\tilde{\bar s},\tilde{\bar b}\bar s)/\Gamma(t\to bW^+)
\simeq 0.16\;(\lambda^{\prime\prime}_{233})^2 \quad
 (for \; m_{\tilde q}=150\;\rm GeV)\;.
\label{rpvsm2}
\end{equation}
These RPV decays would plausibly be followed by $\tilde q\to q\chi^0_1$
and $\chi^0_1\to cbs,\bar c\bar b\bar s$ (via the same
$\lambda^{\prime\prime}_{233}$  coupling with a short lifetime analogous
to Eq.(\ref{chilife})), giving altogether
\begin{equation}
t\to (b\tilde s,s\tilde b)\to bs\chi^0_1 \to (cbbbs,\bar c\bar bbb\bar s).
\label{tcbbbs}
\end{equation}
This all-hadronic mode could in principle be identified experimentally,
through the multiple $b$-jets plus the $t\to 5$-jet and
$\chi^0_1\to3$-jet invariant mass constraints.  However, it would not be
readily mistaken for the SM hadronic mode $t\to bW\to 3$-jet, and would
simply reduce all the SM top signal rates. If the coupling approached
the fixed-point value $\lambda^{\prime\prime}_{233}\simeq 1.0$,
while $m_{\tilde q}\simeq 150$ GeV as assumed in Eq.(\ref{rpvsm2}),
the SM top signals would be suppressed by a factor
$(1+ 0.16(\lambda^{\prime\prime}_{233})^2)^{-2}\simeq 0.75$,
which is strongly disfavored by the present data~\cite{cdf,dzero} but
perhaps not yet firmly excluded.

If indeed the $s$- and $b$-squarks were lighter than $t$ to allow the
$B$-violating modes above, it is quite likely that the $R$-conserving decay
$t\to\tilde t\chi^0_1$ would also be allowed, followed by
$\tilde t\to c\chi^0_1$ (via a loop) and $B$-violating decays for both
neutralinos, with net effect
\begin{equation}
t\to\tilde t\chi^0_1\to c\chi^0_1\chi^0_1 \to
(cccbbbb,ccbb\bar c\bar b\bar b, c\bar c\bar c\bar b\bar b\bar b\bar b).
\end{equation}
This seven-quark mode would look quite unlike the usual SM
modes and would further suppress the SM signal rates.
Depending on details of the sparticle spectrum, however,
other decays such as $\tilde t\to bW\chi^0_1$ might take part too,
leading to different final states; no general statement can be
made except that they too would dilute the SM signals and
therefore cannot be very important.

\section{Conclusions}

The renormalization group evolution of the Standard Yukawa couplings can
be affected by the presence of RPV couplings. In this paper we have done the
following:

\begin{itemize}

\item We have identified the fixed points that occur in the RPV couplings,
under the usual assumption that only $B$-violating or only $L$-violating
RPV interactions exist.

\item These fixed points provide process-independent
upper bounds on RPV couplings at the electroweak scale; we confirm
previously obtained bounds in the $B$-violating case and provide new
results for the $L$-violating case [Fig.1].

\item We have also addressed scenarios with large $\tan \beta $ where
$\lambda_b $ too can reach a fixed point.

\item The fixed point values are summarized in Table~II.  It is interesting
that they are compatible with all present experimental constraints.

\item However, fixed-point values of the $L$-violating coupling
$\lambda^\prime_{333}$ or the $B$-violating coupling
$\lambda^{\prime\prime}_{233}$ would require the corresponding
sparticles to have mass $\agt m_t$ to prevent unacceptably large
fractions of top decay to sleptons or squarks.

\item The fixed points lead to constraints, correlating the RPV couplings
with the top quark Yukawa coupling at the GUT scale, from lower bounds
on the top mass [Fig.2].

\item We have derived evolution equations for the CKM matrix and examined
the evolution of the CKM mixing angles in the presence of RPV
couplings [Fig.3]. In the
most general case, new CKM-like angles occur in the RPV coupling sector
and influence the scaling of the CKM unitarity triangle.

\end{itemize}

\section*{Acknowledgements}
VB thanks Herbi Dreiner for a discussion and the Institute
for Theoretical Physics at the University of California, Santa Barbara for
hospitality during part of this work. RJNP thanks the University of
Wisconsin for hospitality at the start of this study.
This research was supported in part by
the U.S.~Department of Energy under Grant Nos.~DE-FG02-95ER40896 and
DE-FG02-91ER40661, in part by the National Science Foundation under Grant
No.~PHY94-07194, and in part by the University of Wisconsin
Research Committee with funds granted by the Wisconsin Alumni Research
Foundation and support by NSF.
TW is supported by the {\em Deutsche Forschungsgemeinschaft }(DFG).

\end{document}